\documentclass[aps,prb,showpacs,twocolumn,superscriptaddress,ams,%
floatfix]{revtex4}
\usepackage{graphicx}
\usepackage{amsmath}
\usepackage{psfrag}

\def\nn{\nonumber}

\begin{document} 

\title{Fractional Flux Periodicity in Doped Carbon Nanotubes}

\author{K. Sasaki}
\email[E-mail: ]{sasaken@imr.tohoku.ac.jp}
\affiliation{Institute for Materials Research, Tohoku University, 
Sendai 980-8577, Japan}
\author{S. Murakami}
\affiliation{Department of Applied Physics, University of Tokyo, Hongo,
Bunkyo-ku, Tokyo 113-8656, Japan}
\author{R. Saito}
\affiliation{Department of Physics, Tohoku University and CREST, JST,
Sendai 980-8578, Japan}

\date{\today}

\begin{abstract}
 An anomalous magnetic flux periodicity of the ground state is predicted
 in two-dimensional cylindrical surface composed of square and honeycomb
 lattice. 
 The ground state and persistent currents exhibit an approximate
 fractional period of the flux quantum for a specific Fermi energy.
 The period depends on the aspect ratio of the cylinder and on the
 lattice structure around the axis.
 We discuss possibility of this nontrivial periodicity in a heavily
 doped armchair carbon nanotube. 
\end{abstract}

\maketitle

In the Aharonov-Bohm (AB) effect, the phase of wavefunctions is
modulated by a magnetic field, thereby manifesting the quantum nature of
electrons. 
One of the direct consequences of the AB effect in solid state physics
is the persistent current in a mesoscopic
ring~\cite{BIL,Buttiker,Webb,Imry}.
The persistent current is an equilibrium current driven by a magnetic
field threading the ring, and it generally shows the flux periodicity of
$\Phi_0=hc/e$. 
On the other hand, there exist some systems where the period becomes a
general fraction of $\Phi_{0}$, revealing an interference effect between
channels.
At present, theoretical investigations of the fractional flux
periodicity has been done in a two-dimensional (2D) system composed
of a square lattice and in a one-dimensional (1D) system. 
Cheung et al.~\cite{Cheung} found that a finite length cylinder with a
specific aspect ratio exhibits the fractional flux periodicity in the
persistent currents.
The same configuration with an additional magnetic field perpendicular
to the cylindrical surface was analyzed by Choi et al~\cite{Choi}.
They reported that a fractional flux periodicity appears in the
persistent currents. 
Its period is mainly determined by the additional perpendicular flux,
but is also dependent on the number of lattice sites along the
circumference.
We have shown in the previous paper that torus geometry exhibits the
fractional flux periodicity depending on the twist around the torus axis
and the aspect ratio~\cite{SKS}.
As for 1D systems, 
Kusmartsev et al.~\cite{Kusmartsev} reported on
fractional AB effect in a certain limit of the Hubbard model and Jagla
et al.~\cite{Jagla} found that correlations change the fundamental
periodicity of the transmittance of an AB ring with two contacts.
In 1D, there is only one channel and therefore there exists no
coherence effect between channels; hence the origin of the fractional
flux periodicity differs from that of two-dimensional system.

However, since these geometries have not yet been realized
experimentally, it is important to examine if existing materials with
cylindrical geometry can be used to detect such an interference effect.
In this paper, we will show that an approximate fractional flux
periodicity appears in a honeycomb lattice cylinder, which is realized
in a single wall carbon nanotube (SWNT)~\cite{SDD}. 
The fractional periodicity requires a heavy doping, shifting the Fermi
energy up to the energy of the transfer integral (2.9 eV).
The persistent current in doped nanotubes was also theoretically
examined by Szopa et al.~\cite{Szopa}, though they do not mention the
fractional periodicity. 

\begin{figure}[htbp]
 \begin{center}
  \psfrag{F}{$\Phi$}
  \psfrag{s}{periodic}
  \psfrag{N}{$N$}
  \psfrag{M}{$M$}
  \psfrag{T}{$l_T |T|$}
  \psfrag{C}{$|C_h|$}
  \psfrag{I}{$I_{\rm pc}(N_\Phi)$}
  \includegraphics[scale=0.35]{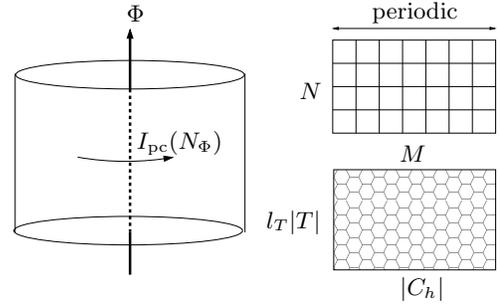}
 \end{center}
 \caption{(left) Geometry of a 2D cylinder in the presence
 of the AB flux, $\Phi$.
 (right) We consider 2D cylinders composed of a square
 and an armchair nanotube.}
 \label{fig:cylinder}
\end{figure}
Here we consider the interference effect in a 2D cylinder
(Fig.~\ref{fig:cylinder}) composed 
of a honeycomb lattice, and calculate the ground state energy and the
persistent current.
The persistent
current ($I_{\rm pc}(N_\Phi)$) is defined by
differentiating the ground state energy
($E_0(N_\Phi)$) with respect to the magnetic flux
penetrating through a hollow core of the cylinder
(See Fig.~\ref{fig:cylinder}) as $I_{\rm
pc}(N_\Phi) \equiv -\partial E_0(N_\Phi)/ \partial
\Phi$.
We show that the fraction of flux period ($\Phi_0/Z$) depends on the
aspect ratio of a cylinder: $1/Z = n/l_T$ for a doped $(n,n)$ armchair
SWNT with $l_T$ being the number of unit cells along the cylindrical
axis.
Persistent current can be observed experimentally via the induced
magnetic moment of the system, which was recently demonstrated in a SWNT
by Minot et al~\cite{Minot}.

First we consider a finite length 2D cylinder composed of a square
lattice, as studied by Cheung et al~\cite{Cheung}. 
The length of the cylinder and the circumferential length are $Na$ and
$Ma$, respectively, where $a$ is the lattice constant.
By solving the nearest-neighbor tight-binding Hamiltonian with hopping
integral $t$, we obtain the energy eigenvalue as 
\begin{align}
 E_{nm} &= -2t \left\{ \cos \left( \frac{n\pi}{N+1} \right) 
 + \cos \left( \frac{2\pi m}{M} \right) \right\}, \nn \\
 & \text{where $1\le n \le N$ and $-\frac{M}{2}+1 \le m \le \frac{M}{2}$}.
 \label{eq:2d-eigen}
\end{align}
In the presence of the AB flux parallel to the 
cylinder axis, $E_{nm}$ changes according to the gauge
coupling which can be obtained by substituting $m \to m-N_\Phi$ where
$N_\Phi$ is the number of flux, defined by $N_\Phi \equiv \Phi/\Phi_0$
(see Fig.~\ref{fig:cylinder}). 

When we consider a half-filling system ($E_{\rm F}=0$),
the ground state energy $E_0(N_\Phi)$ is given by
\begin{align}
 E_0(N_\Phi) = \sum_{n=1}^N
 \sum_{m=\left[ - A_n + N_\Phi \right]+1}^{\left[ A_n + N_\Phi \right]}
E_{nm}(N_\Phi),
 \label{eq:vac-ene}
\end{align}
where $A_n \equiv \frac{M}{2} \left( 1- \frac{n}{N+1} \right)$ and
$[x]$ represents the largest integer smaller than $x$.  From
Eq.~(\ref{eq:vac-ene}), the persistent current is expressed as
\begin{align}
 I_{\rm pc}(N_\Phi) = \sum_{n=1}^{N} 
 \sum_{m=\left[ - A_n + N_\Phi \right]+1}^{\left[ A_n + N_\Phi \right]}
 \frac{2et}{M} \sin \left( \frac{2\pi(m-N_\Phi)}{M} \right).
 \label{eq:pc-explicit}
\end{align}
For $M \gg 1$, we can rewrite Eq.~(\ref{eq:pc-explicit}) as 
\begin{align}
 \frac{I_{\rm pc}(N_\Phi)}{I_0} =
 \frac{2}{M} \sum_{j=1}^\infty C_j \frac{\sin 2\pi N_\Phi j}{j},
 \label{eq:pc-2dim-final}
\end{align}
where $I_0 = 2et/ (M\sin \frac{\pi}{M})$ and we ignored the correction
of ${\cal O}(M^{-2})$ to the right hand side. 
The coefficient, $C_j$ is given by  
\begin{align}
 C_j
 \equiv \sum_{n=1}^N 
 \sin \left( \frac{2\pi A_n}{M} \right) \cos \left( 2\pi j A_n
 \right).
 \label{eq:C_j}
\end{align}
When the aspect ratio satisfies $M/2(N+1) = 1/Z$ with integer $Z$, 
$\cos (2\pi Z A_n) = 1$ holds for all $n$ and $C_{j}$ satisfies $C_{j+Z}
= C_j$. 
Furthermore, in the limit of $N \gg 1$, $C_Z$ becomes large compared
with $C_1,\cdots,C_{Z-1}$ since they are proportional to
${\cal O}(1/N)$ and $C_Z \propto {\cal O}(N)$.
Since only $j=Z,2Z,\cdots$ are dominant in Eq.~(\ref{eq:pc-2dim-final}),
we get an approximate fractional ($\Phi_0/Z$) periodicity of $I_{\rm
pc}$ when $M/2(N+1) = 1/Z$ which becomes exact in the long length limit
$N \gg 1$, as was found by Cheung et al.~\cite{Cheung}

Let us apply this result to doped armchair SWNT.
The carbon nanotubes can be specified by the chiral vector,
$\mathbf{C}_{h} = n \mathbf{a}_1 + m \mathbf{a}_2$, and the
translational vector, $\mathbf{T} = t_1 \mathbf{a}_1 + t_2
\mathbf{a}_2$, where $(n,m)$ and $(t_1,t_2)$ are integers and
$\mathbf{a}_1,\mathbf{a}_2$ are symmetry translational vectors on the
planar honeycomb lattice\cite{SDD} with
$|\mathbf{a}_{1}|^{2}=|\mathbf{a}_2|^{2}
=2\mathbf{a}_{1}\cdot\mathbf{a}_{2}$.
The chiral vector $\mathbf{C}_{h}$ specifies the circumference of the
cylinder, and the unit cell of the nanotube is defined by two mutually
perpendicular vectors $\mathbf{C}_{h}$ and $\mathbf{T}$.
The length of the cylinder is specified by a vector $l_T \mathbf{T}$. 
We decompose the wave vector as $\mathbf{K} = \mu_1 \mathbf{K}_1
+ (\mu_2/l_T) \mathbf{K}_2$ where $\mu_1$ and $\mu_2$ are integers, and
the wave vectors $\mathbf{K}_1$ and $\mathbf{K}_2$ are defined by the
condition~\cite{SDD}: $\mathbf{C}_{h} \cdot \mathbf{K}_1= 2\pi$,
$\mathbf{C}_{h} \cdot \mathbf{K}_2 = 0$, $\mathbf{T} \cdot \mathbf{K}_1
= 0$ and $\mathbf{T} \cdot \mathbf{K}_2 = 2\pi$. 
With these definitions, the energy eigenvalue of the valence electrons
can be expressed by
\begin{align}
 E_{\mu_1 \mu_2}(N_\Phi)
 = -V_\pi \sqrt{1+4\cos X \cos Y + 4\cos^2 Y},
\end{align}
where $V_\pi (=2.9{\rm eV})$  is the transfer integral for
nearest-neighbor carbon atoms and the variables $X$ and $Y$ are defined
by
\begin{align}
 & X = \frac{\pi}{N_c} \left( -(t_1-t_2)(\mu_1 - N_\Phi) + (n-m) 
 \frac{\mu_2}{l_T} \right), \\ 
 & Y = \frac{\pi}{N_c} \left( -(t_1+t_2)(\mu_1 - N_\Phi) + (n+m) 
 \frac{\mu_2}{l_T} \right),
\end{align}
where $N_c = mt_1 - nt_2$.
This implies that the motions around the axis ($\mu_1$) and along the
axis ($\mu_2$) couple with each other.

Without doping (i.e., at half-filling), the energy dispersion relation
of carbon nanotube has only two distinct Fermi points called $K$ and
$K'$ Fermi points.
In this case only two channels touch the Fermi level and the ground state
energy shows the AB effect with the periodicity of
$\Phi_0$~\cite{Ando,Roche} and corresponding situation has reported
experimentally~\cite{Kono,Minot}. 
We consider a special doping which shifts the Fermi level from $0$ to
$E_F = - V_\pi$~\cite{Szopa} (the case of $E_F=V_\pi$ can be analyzed in
the same way and the results are the same because of the particle-hole
symmetry).
The energy dispersion relation around the Fermi level is
approximately given by 
\begin{align}
 E_{\mu_1 \mu_2} 
 \approx E_F - 2V_\pi \cos Y \left( \cos X + \cos Y \right).
 \label{eq:cos}
\end{align}
We note that the energy dispersion relation is
similar to that of the square lattice except for
the factor of $\cos Y$.  
Here, we adopt the armchair SWNTs ($n=m$) because we can set $t_1=1$ and
$t_2=-1$ and therefore $X$ is proportional to only $\mu_1$ and $Y$ to
$\mu_2$, which results in that $\cos X + \cos Y$ becomes the energy
dispersion relation of a square lattice.

\begin{figure}[htbp]
 \begin{center}
  \psfrag{x}{$N_\Phi$}
  \psfrag{y}{Ground state Energy $[V_\pi]$}
  \psfrag{a}{(a)}
  \psfrag{b}{(b)}
  \includegraphics[scale=0.6]{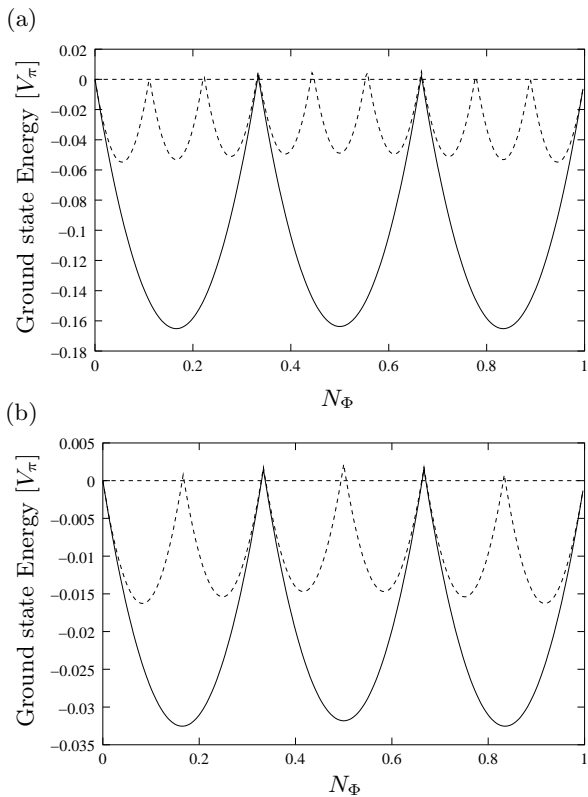}
 \end{center}
 \caption{Ground state energy 
as a function of the flux for $(n,n)$ armchair SWNT.
(a) $n=30$ (diameter $\sim $ 4 nm) with two
 different lengths: $l_T/n = 3$ (length $\sim$ 22 nm) (solid curve), 
 $l_T/n = 9$ (dashed curve), and 
 (b) $n=80$ (diameter $\sim$ 10 nm)
 with two different lengths: $l_T/n = 0.6$ (length 
$\sim$ 11.6 nm) (solid curve), $l_T/n = 1.2$ (dashed curve).
We offset the origin of the energy for comparison.}
 \label{fig:ground-ene}
\end{figure}

We first show the numerical results of the ground state energy
$E_0(N_\Phi)$ for the doped armchair SWNT. 
In Fig.~\ref{fig:ground-ene}(a), we plot $E_0(N_\Phi)$ of doped armchair 
$(30,30)$ SWNTs with different lengths.
The ground state energy shows an approximate fractional flux periodicity
depending on the aspect ratio defined by $l_T/n$.
We can see that the approximate fractional periodicity ($\Phi_0/Z$) corresponds
to $Z = l_T/n$. 
However, there also exists the approximate fractional flux periodicity for a 
shorter nanotube, $l_T < n$~\cite{SKS}. 
In Fig.~\ref{fig:ground-ene}(b), we plot the ground state energy for
$(80,80)$ armchair SWNT with $l_T = 48$ ($l_T |T|\approx 11.6 [{\rm
nm}]$) and $l_T=96$ ($l_T |T| \approx 23.2 [{\rm nm}]$). 
The approximate fractional periodicity corresponds to $1/Z = 1/3$ and
$1/6$, respectively, in which $l_T/n$ is $3/5$ and $6/5$. 

\begin{figure}[htbp]
 \begin{center}
  \psfrag{X}{$X$}
  \psfrag{Y}{$Y$}
  \psfrag{p}{$\pi$}
  \psfrag{a}{(a)}
  \psfrag{b}{(b)}
  \includegraphics[scale=0.35]{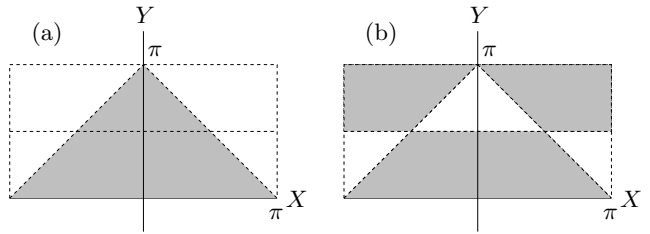}
 \end{center}
 \caption{Occupied states are shown in gray color in the Brillouin zone
 for the energy dispersion relations;
 (a) $-2V_\pi |\cos Y| (\cos X + \cos Y)$ in Eq.~(\ref{eq:p-cos}) and 
 (b) $-2V_\pi \cos Y (\cos X + \cos Y)$ in Eq.~(\ref{eq:cos}).}
 \label{fig:Bzone}
\end{figure}

Now we derive the fractional flux periodicity of persistent current in
doped armchair SWNTs analytically with use of the result for the square
lattice.
Equation (\ref{eq:2d-eigen}) for the square lattice is similar to
Eq.~(\ref{eq:cos}) for the doped SWNT.
Although there remains a factor of $\cos Y$ difference, it does not
affect the fractional periodicity; we can analytically 
show below that the fractional periodicity $\Phi_{0}/Z$ occurs when
$l_T/n=Z$ is an integer.  
To see this, it is useful to consider a hypothetical energy dispersion,
\begin{align}
 -2V_\pi |\cos Y| (\cos X + \cos Y).
 \label{eq:p-cos}
\end{align}
This energy eigenvalue corresponds to multiplying the factor $\left| \cos
n\pi/(N+1) \right|$ to the $E_{nm}(N_\Phi)$ for the square lattice. 
The corresponding persistent current can be expressed by
multiplying $C_j$ in Eq.~(\ref{eq:pc-2dim-final}) by the factor of
$\left| \cos n\pi/(N+1) \right|$,
\begin{align}
 C_j
 = \sum_{n=1}^N \left| \cos \frac{n\pi}{N+1} \right|
 \sin \left( \frac{2\pi A_n}{M} \right) \cos \left( 2\pi j A_n
 \right).
 \label{eq:C_j_new}
\end{align}
Thus the persistent current for Eq.~(\ref{eq:p-cos}) 
still exhibits the fractional periodicity
depending on the aspect ratio.
To compare Eq.~(\ref{eq:cos}) with
Eq.~(\ref{eq:p-cos}), 
we consider the difference between the ground state energies for the two
energy dispersions.
Figure \ref{fig:Bzone} shows the occupied states (gray color) for (a)
the hypothetical energy dispersion of Eq.~(\ref{eq:p-cos}) and (b)
that for doped SWNT.
From this figure, it follows that the difference of ground 
state energies is a constant independent of $N_\Phi$, because it
corresponds to the sum of energy of all states for dispersion
Eq.~(\ref{eq:p-cos}) within the region of $-\pi\leq X\leq\pi$ and
$\pi/2\leq Y\leq \pi$. 
This shows that the persistent current in the doped armchair SWNTs 
has the same periodicity as the square-lattice cylinder after the 
replacement $t \to V_\pi$, $M \to 2n$ and $N+1 \to l_T$;
the period is thus given by $\Phi_{0}/Z$ for 
an integer value of $Z=l_T/n$.
Moreover, even when $l_T/n$ is equal to a rational number $Z/q$ for 
coprime integers $Z$ and $q$, such as $l_T/n = 3/5$,
the system shows an approximate fractional flux periodicity 
$\Phi_0/Z$.
It is because in Eqs.~(\ref{eq:C_j}) and (\ref{eq:C_j_new}) $C_{Z},C_{2
Z},\cdots$ are much larger than other $C_j$, owing to $\cos (2\pi Z
A_n)=1$.

Let us comment on the magnetic field $B_n$
for $(n,n)$ armchair SWNT which corresponds to
$\Phi_0 = 4 \times 10^{-7} [{\rm gauss} \cdot {\rm
cm}^2]$. For $n=100$ (diameter is about
13 nm), $B_n$ is about 30T, which is experimentally attainable.
For the fractional
periodicity, the period becomes 
$B_n/Z\sim 30{\rm T}/Z$ , which is easily achieved in 
experiments for larger $Z$.
Although $Z$ can be large in the long length limit
of an armchair SWNT ($l_T \gg n$), the
periodic motion of the electron along the axis of
long nanotubes may be affected by 
decoherence effect such as lattice
deformations or defects.  It is
also valuable to comment on the magnetic moment ($\mu_{\rm orbit}$) 
of the system~\cite{Minot}.  The magnetic moment can
be calculated directly from the persistent current
as $\mu_{\rm orbit} = S I_{\rm pc}(N_\Phi)$, where
$S$ is the cross-sectional area of the tube ($S =
|\mathbf{C}_{h}|^2/4\pi$).  When $l_T/n=Z$ ($Z$ is an
integer), we estimate that the maximum amplitude
of the magnetic moment scales as $\mu_{\rm
orbit}/\mu_B = {\cal O}(1) n^2$, where $\mu_B = 6
\times 10^{-2} [{\rm meV}/{\rm T}]$ is the
Bohr magneton.

The Fermi energy ($E_F = \pm V_\pi\sim \pm 2.9{\rm eV}$) for doped
SWNT for the fractional flux periodicity
might be difficult for chemical doping, whereas 
it might be possible for electrochemical doping;
a recent experiment of electro-chemical gating
achieves Fermi energy shift of order $\pm 1{\rm eV}$ \cite{Kruger}.

\begin{figure}[htbp]
 \begin{center}
  \psfrag{x}{$N_\Phi$}
  \psfrag{y}{Ground state Energy $[V_\pi]$}
  \includegraphics[scale=0.6]{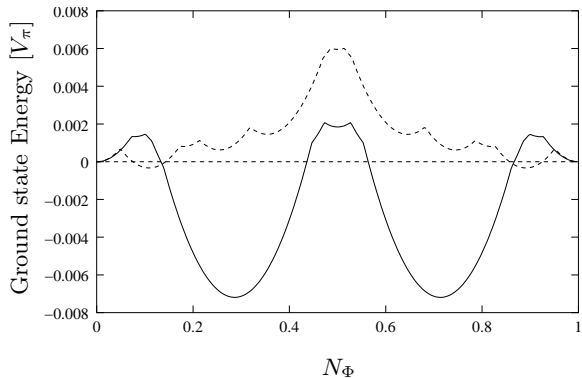}
 \end{center}
 \caption{Ground state energy of a weak doped ($E_F = - V_\pi/10$)
 chiral nanotubes. 
 We consider $(60,30)$ chiral nanotube with $l_T = 15$ (solid curve) and
 $(70,30)$ with $l_T = 5$ (dashed curve). 
The origin of the energy has been shifted to facilitate comparison.
 }
 \label{fig:ground-ene-weak}
\end{figure}

Finally we point out that the fractional nature can be seen even for the 
smaller Fermi energy. 
In Fig.~\ref{fig:ground-ene-weak} we plot $E_0(N_\Phi)$ with a weak
doping ($E_F = - V_\pi/10$) for for $(60,30)$ (solid curve) and
$(70,30)$ (dashed curve) chiral nanotube with $l_T = 15$ and $l_T = 5$, 
respectively.  
These curves clearly show some coherence effects.
Thus a reproducible coherence pattern of the magnetic moment may be
obtained as a function of magnetic field for doped SWNTs. 

It is not easy to obtain of a SWNT with the diameter of 4nm. 
Thus we generally need a multi-wall carbon nanotube in which the current
flows only few outermost layers. 
In this case we need to consider an effect of a finite thickness of
layers. 
If the periodicity comes from the some interlayer coupling, complex
coherence effect would appear. 
In fact, there are several experimental reports on an approximately
fractional flux periodicity in
magneto-resistance~\cite{Bachtold,Fujiwara,AAS}.   
Bachtold et al.~\cite{Bachtold} observed an oscillation with a period
$\sim \Phi_{0}/10$.
It not be explained by the Al'tshuler-Aronov-Spivak theory which
predicts the period of $\Phi_{0}/2$~\cite{Bachtold,AAS}. 
At present moment, our results do not account for these oscillations,
which will be a future work.

In conclusion, we calculated numerically and analytically the
fractional periodicity of the ground state energy and persistent
currents in 2D cylinders composed of a square and doped armchair SWNT.
A doped armchair SWNT
also exhibits the fractional periodicity when they are doped to $E_F =
\pm V_\pi$.  The fraction ($\Phi_0/Z$) depends on the aspect ratio
given by $Z = l_T/n$,
where $l_T$ is the number of unit cells along the cylindrical axis.  
An experimental
investigation of the AB effect in a doped SWNT gives a key
to understand this special coherence phenomenon.

\begin{acknowledgments}
 K. S. is supported by a fellowship of the 21st Century COE Program of
 the International Center of Research and Education for Materials of
 Tohoku University. 
 S. M. is supported by Grant-in-Aid (No.~16740167) from the Ministry of
 Education, Culture, Sports, Science and Technology, Japan.
 R. S. acknowledges a Grant-in-Aid (Nos.~13440091 and 16076201) from the
 Ministry of Education, Culture, Sports, Science and Technology, Japan.
\end{acknowledgments}


\end{document}